\documentclass[aps,preprint,eqsecnum,prl]{revtex4}
\usepackage{epsfig}

\newcommand{\beq}{\begin{equation}}
\newcommand{\eeq}{\end{equation}}

\newcommand{\vev}{{\langle\, \phi\, \rangle}}

\newcommand{\remove}[1]{}
\newcommand {\comment}[1]{}             
\begin{document}
%{\tighten
%\preprint{\vbox{\hbox{EUPC/00--01}}}

%\baselineskip 16pt
%\renewcommand\baselinestretch{1.2}

\title{Twisted Alice Loops as Monopoles}

\author{Katherine M.~Benson}
\email{benson@physics.emory.edu}
\affiliation{
Department of Physics,
Emory University,  400 Dowman Drive, Suite N202, Atlanta, GA\ 30322}
\author{Tom Imbo}\email{imbo@uic.edu}
\affiliation{Department of Physics,
University of Illinois at Chicago,
845 W. Taylor St,  m/c 273,
Chicago, IL 60607-7059}

%\bigskip
\date{\today}

\begin{abstract}
Symmetry breaking can produce ``Alice'' strings, which alter scattered
charges and carry monopole number and charge when twisted into loops.
We apply recent topological results, fixing Alice strings' stability
and prescribing their twisting into loops with monopole charge, to
several models.  We show that Alice strings of condensed matter
systems (nematic liquid crystals, $^3$He-A, and related systems of
non-chiral Bose condensates and amorphous chiral superconductors) are
topologically Alice, and carry fundamental monopole charge when
twisted into loops. They might thus be observed indirectly, not as
strings, but as loop-like point defects. Other
models yield Alice loops that carry only deposited, and not
fundamental, charge.

\end{abstract}

\maketitle

%\renewcommand{\baselinestretch}{1.6}

%\pacs{11.27.+d, 11.30.Fs, 11.30.Ly,98.80.Cq}
% end the tighten
%\narrowtext
%\newpage

%\renewcommand{\baselinestretch}{1.6}
\section{Introduction}

Among the defects created in spontaneous symmetry breakdown are Alice
strings.\cite{oldAlice,stringzm,newAlice} Like monopoles
\cite{moncolor}, Alice strings obstruct the global extension of
unbroken symmetries, making them multivalued when parallel transported
around the string. This algebraic obstruction causes nonconservation
of associated charges, when Aharonov-Bohm scattered around the string;
it also induces monopoles, as twisted loops of Alice string. Alice
strings arise in both particle physics and condensed matter models,
with condensed matter systems offering the most likely prospects for
their observation. \cite{volbook} We show here that known condensed
matter Alice strings form twisted loops with fundamental monopole
charge, suggesting a second avenue for their potential
observation: Alice strings might be observed, not as strings, but as looplike
point defects, when twisted loops comprise the energetically favored
solution of fundamental monopole charge.

We recently established in \cite{topalice} a topological
criterion for Alice behavior, stating when Alice strings {\bf must}
form, and when strings' Alice features may be deformed away. Consider the symmetry breakdown of Lie group $G \rightarrow H$, taking for $G$ the simply connected cover of the initial Lie symmetry. A
topological string then has homotopy
$\pi_o(H)$; that is, its flux $U(2\pi)$ lies in a disconnected component
of the unbroken symmetry group $H$.  Monopoles have homotopy $\pi_1(H)$, describing loops
$h(\alpha)$ of different winding in $H$. Our criterion labels strings
with flux $U(2\pi)$ topologically Alice if they alter the topological
charge of monopoles circumnavigating them:
$$h(\alpha) \rightarrow \tilde{h} (\alpha) = U(2\pi)\ h(\alpha)\
U^{-1}(2\pi)\ \not\sim \ \ h(\alpha) \ .$$ This is a topological criterion,  corresponding to a nontrivial action of $\pi_o(H)$ on $\pi_1(H)$, where
$h_o = U(2\pi) \in \pi_o(H)$ alters the topological winding of loop
$h(\alpha) \in\pi_1(H)$: \beq \tilde{h}
(\alpha) = h_o\ h(\alpha)\ h_o^{-1} \ \ \not\sim \ \ h(\alpha)\ \
.\label{crit}\eeq

Using this criterion we constructed a prescription for twisting
Alice strings into loops carrying monopole charge. We showed that the
twisted Wilson line
\beq U(\varphi, \alpha) = h^{-1}(\alpha)\ U(\varphi)\ h(\alpha)\
h_o^{-1}\ ,\label{twistWil}\eeq
with $h(\alpha)$ and $h_o$ as in our criterion above,
generates a single-valued twisted Alice loop.  $U(\varphi,\alpha)$
interpolates between the point $h_o^{-1}$ at $\varphi = 0$ and the
loop $h^{-1}(\alpha)\ \tilde{h}(\alpha)$ at $\varphi = 2\pi$. The
twisted Alice loop thus carries a nontrivial monopole charge $
h^{-1}(\alpha)\ \tilde{h}(\alpha)$ if and only if the Alice string
obeys our topological criterion. This monopole charge corresponds to
that deposited on the Alice loop in the monopole circumnavigation
$h^{-1}(\alpha) \ \rightarrow \
\tilde{h}^{-1}(\alpha)$. \cite{topalice}

Two points are key. First, our topological arguments only indicate
that {\bf deposited} monopole charge can be carried by a twisted Alice
loop. Monopole charge is typically not deposited in single fundamental
units, leaving open the question of whether twisted Alice loops can
carry {\bf fundamental} monopole charge. Second, we took as twisting
function $h(\alpha)$ for the twisted Wilson line $U(\varphi, \alpha)$,
the loop in $H$ representing a fundamental monopole (call this the
fully-twisted Alice loop). We remain free to choose a different
twisting function $h(\alpha)$ in $H$, so long as it renders
$U(\varphi, \alpha)$ singlevalued in the angle $\alpha$. Propitious
choice of $h(\alpha)$, in some models, allows construction of twisted
Alice loops with {fundamental} monopole charge.

We exploit both points here, in examining the monopole charge carried
by model twisted Alice loops. We study primarily those models which
offer the best condensed matter candidates for Alice strings: the
original Alice string of Schwarz, coinciding with the Alice string of
liquid crystals and of non-chiral Bose condensates \cite{oldAlice,
leonvol}; and the Alice string of $^3$He-A \cite{volmin}, arising anew
for unconventional spin-triplet superconductors \cite{spintrips}. In
both models we find strings which are topologically Alice. We also
find twisted Alice loops supporting fundamental monopole charge,
even though monopole charge deposits onto string loops in even
increments.  Thus, for the Alice strings of interest to condensed
matter, even the most fundamental singular point defect, or monopole,
can take the form of a twisted Alice loop.

We show, in other models, different outcomes for the topology of Alice
strings and their twisted loops. When Alice behavior is
nontopological, the fully-twisted Alice loop has trivial monopole
charge. However, an Alice loop with different twist $h(\alpha)$ may
still exist, carrying nontrivial monopole charge. We examine these
possibilities in the context of a nontopologically Alice string
discussed in \cite{stringzm} and \cite{topalice}. Finally, string
loops which are topologically Alice may admit {\bf only} the full
twisting invoked in our topological argument. In this case, twisted
Alice strings carry {\bf only} deposited, and not fundamental,
monopole charge. We realize this possibility for a new topological
Alice string, formed in the symmetry breakdown $SU(3) \rightarrow
O(2)$. Here twisted Alice loops support only even
deposited monopole charge, and cannot form fundamental monopoles.

Through these models we show that whether twisted Alice loops can
support fundamental monopole charge depends closely on the
symmetry-breaking pattern.  Specifically, it depends on the initial
symmetry group $G$, through the identification of loops in $H$ with
monopoles via the exact sequence for $\pi_2(G/H)$. When fundamental
monopoles correspond to nonminimal-winding loops in $H$, alternative
choices for the loop twisting $h(\alpha)$ exist; and these can induce
a twisted Alice loop with fundamental monopole charge.  Algebraic
commutations in the model may also allow alternative twistings
$h(\alpha)$, yielding twisted Alice loops with fundamental monopole
charge.

\section{\label{model:canon}The Schwarz, or Nematic,  Alice String}

We start with the simplest example, the canonical Schwarz Alice
string, \cite{oldAlice} whose symmetry-breaking pattern coincides with
Alice strings in nematic liquid crystals and in non-chiral Bose
condensates. \cite{leonvol}

Here $G$ is $SO(3)$, with
Higgs $\phi$ transforming in the adjoint representation.
When $\phi$ develops the vev
$\vev = {\rm diag}\ (1,1,-2)\ \ ,
$
$SO(3)$ breaks to the residual symmetry $H = O(2)$, containing  z-rotations $R_z\,(\alpha)$ and the discrete symmetry element
$ h_o = R_x\, (\pi) = {\rm diag}\ (1, -1, -1)\ .$
Here $\pi_o (H) = Z_2$ and $\pi_1(H) = Z$ so we have topological strings and monopoles. The canonical Alice string has Wilson line
$ U(\varphi) = R_x\, (\varphi/2)$
with $U(2\pi) = h_o$. This string is Alice, as $U(2\pi)$ fails to commute
with the unbroken symmetry generator $T_z$; in fact, on parallel
transport around the string,
\beq T_z \rightarrow U\ (2\pi) \ \ T_z \ \ U^{-1}\ (2\pi) = -T_z \ \ .
\label{canAliceT}\eeq

This canonical Alice string meets our topological criterion, of changing
topological monopole charge upon circumnavigation. By the exact
sequence for $\pi_2 (G/H)$, topological monopoles are associated with
nontrivial loops in $O(2)$ which can be unwound in $SO(3)$. Since only
even winding loops in $O(2)$ can be unwound in $SO(3)$, the
fundamental monopole in this canonical Alice model has a loop in
$O(2)$ of winding 2.

In applying our topological criterion, we choose $h_o$ as our
representative of the string, a nontrivial element of
$\pi_o (H)$, and $h(\alpha) = R_z(2\alpha)$ as our representative of
the fundamental Alice monopole, a winding 2 element of $\pi_1(H)$. This gives
$$\tilde{h} (\alpha) = h_o\ h(\alpha)\ h_o^{-1}  = h^{-1} \, (\alpha)\ \ ,$$
from equation (\ref{canAliceT}). Note that $h^{-1} \, (\alpha)$ has $O(2)$ winding -2, topologically distinct from $h(\alpha)$ of $O(2)$ winding 2. Thus
$ \tilde{h} (\alpha)\ \not\sim \
h(\alpha)$ and our topological criterion is fulfilled.

We now construct a monopole as a twisted Alice loop. From Eq.~(\ref{twistWil}), the twisted Wilson line
$$ U(\varphi, \alpha) = h^{-1}(\alpha/2)\ U(\varphi)\ h(\alpha/2)\ h_o^{-1}\ \
$$ generates an Alice loop with single-valued condensate. (We take
$h(\alpha/2)$ because we need only for $h$ to be single-valued in
$\alpha$, and $h(\alpha/2)$, the winding 1 loop in $O(2)$, first
fulfills that requirement.)  $U(\varphi, \alpha)$ interpolates between
$ h_o^{-1}$ at $\varphi = 0$ and $ h^{-1}(\alpha)$ at $\varphi =
2\pi$. It is thus the fundamental antimonopole in the model. Note that
the inverse twisted Alice loop, with twisted Wilson line $
U^{-1}(\varphi, \alpha)$, generates the fundamental monopole.

\section{\label{model:helium3a}The  Alice String of $^3$He-A}

A more complicated global symmetry-breaking pattern describes the
Alice string expected in $^3$He-A \cite{volmin,volbook}, and more recently
predicted in amorphous chiral superconductors with p-wave pairing,
such as Sr$_2$RuO$_4$.\cite{spintrips}

Here $G$ is $SO(3)_L\times SO(3)_S\times U(1)_N$, describing spatial
rotations, spin rotations, and a $U(1)$ phase symmetry associated with
number conservation of helium atoms. (The U(1) symmetry is
approximate, as is independence of spin and orbital rotations due to
minimal spin-orbit coupling, but both describe $^3$He-A well.)  The
matrix order parameter $A$ transforms under symmetry transformations
as
$A \ \rightarrow e^{2i\theta}\ R_S \ A\ R_L^{-1}\ ,$
where $R_S$ and $R_L$ are spin and orbital rotations, respectively.

The order parameter develops the form
$$A_{ij} = \Delta_A\ \hat{d}_i\ (\hat{m}_j + i\,\hat{n}_j)\ \ ,$$ where
$\hat{m}$ and $\hat{n}$ are perpendicular, determining $\hat{l} =
\hat{m} \times\hat{n},$ the direction of the
condensate's angular momentum vector. This breaks $G$ to the residual symmetry
$H= U(1)_{S_{\hat{d}}} \ \times\ U(1)_{L_{\hat{l}} - N/2}\ \times Z_2,$
consisting of spin rotations about the $\hat{d}$ axis; spatial
rotations about the $\hat{l}$ axis when compensated by a matching
$U(1)_N$ phase rotation, and the discrete $Z_2$ transformation $h_o$, with
$h_o: \hat{d}, \ \ \hat{m} + i\,\hat{n}\ \rightarrow \   - \hat{d}, \ \
 -\ (\hat{m} + i\,\hat{n}) \ .$

Identifying the defect topology requires care in this setting, as the
exact sequences relating $\pi_2(G/H)$ and $\pi_1(G/H)$, the monopole
and string homotopy groups, to $\pi_1(H)$ and $\pi_0(H)$ are highly
nontrivial.  Note that $\pi_2(G/H) = Z$, corresponding to the loops
$\pi_1( U(1)_{S_{\hat{d}}} )$, which can be unwound in $G$. (Loops of
the other $U(1)$ factor cannot be unwound in $G$, as they contain
unshrinkable $U(1)_N$ loops.) $\pi_1(G/H) = Z_4$, which describes
strings of two different origins. First, the Alice strings, called
half-quantum vortices, have Wilson lines ending in a disconnected
component of $H$, getting topological stability from
$\pi_0(H)$. Second, a $Z_2$ winding one vortex, nontrivial in
$SO(3)_L$ in $G$, induces as its image a $Z_2$ winding one vortex in
$G/H$, with topological stability inherited from $\pi_1 (G)$. These
two classes of vortices are not independent: instead winding twice
about a half-quantum vortex is equivalent to once around a winding
one vortex, and the full string homotopy is $\pi_1(G/H) = Z_4$, or
windings $0, \pm 1/2$, and $1$ modulo 2, with Alice strings corresponding to
windings $\pm 1/2$.

The Volovik-Mineev Alice string, of winding $\pm 1/2$, has order
parameter $A_{ij}$ with $\hat{d} = \hat{x}$ in spin space, and
$\{\hat{l}, \hat{m},\hat{n}\} = \{\hat{x}, \hat{y}, \hat{z}\}$ in ordinary
space. This is acted on by Wilson line
$ U(\varphi) =  e^{\pm i\varphi/2}\ R_{S_{\hat{z}}}(\varphi/2)$
to give, asymptotically in $r$,
$$A_{ij} (\varphi) = \Delta_A\  e^{\pm i\varphi/2}\ (\cos (\varphi/2)\ \hat{x}_j + \sin  (\varphi/2)\ \hat{y}_j)_S\ (\hat{x}_j + i\hat{y}_j)_L\ \ ,$$
single-valued in $\varphi$. Note that  $U(2\pi) = - R_{S_{\hat{z}}}(\pi) $ lies in the same homotopy class as $h_o$. This string is Alice, making
unbroken symmetry generator $T_{S_{\hat{x}}}$ double-valued.
Physically, this means that a particle flips its spin, and hence its magnetization, on circumnavigating the Alice string.

This long-studied Alice string meets our topological criterion, of changing
topological monopole charge upon circumnavigation. By the exact
sequence for $\pi_2 (G/H)$, topological monopoles are associated with
nontrivial loops in $ U(1)_{S_{\hat{x}}}$ which can be unwound in $SO(3)_S$. As in the nematic case,  only
even winding loops in $ U(1)_{S_{\hat{x}}}$ can be unwound in $SO(3)_S$. Thus the
fundamental monopole in $^3$He-A corresponds to  a loop in
$ U(1)_{S_{\hat{x}}}$ of winding 2.

In applying our topological criterion, we choose $U(2\pi)$ as our
representative of the string, a nontrivial element of
$\pi_o (H)$, and $h(\alpha) = R_{S_{\hat{x}}}(2\alpha)$ as our representative of
the fundamental monopole, a winding 2 element of $\pi_1( U(1)_{S_{\hat{x}}})$. This gives
$$\tilde{h} (\alpha) = h_o\ h(\alpha)\ h_o^{-1}  = h^{-1} \, (\alpha)\ \ ,$$
since $T_{S_{\hat{x}}}\rightarrow - T_{S_{\hat{x}}}$. Note that $h^{-1} \, (\alpha)$ has $U(1)_{S_{\hat{x}}}$ winding -2, topologically distinct from $h(\alpha)$ of $U(1)_{S_{\hat{x}}}$ with winding 2. Thus
$ \tilde{h} (\alpha)\ \not\sim \
h(\alpha)$, meeting our topological criterion.

As in the nematic case, we construct a monopole as a twisted Alice loop. From Eq.~(\ref{twistWil}), the twisted Wilson line
$$ U(\varphi, \alpha) = h^{-1}(\alpha/2)\ U(\varphi)\ h(\alpha/2)\ h_o^{-1}\ \
$$
generates an Alice loop with single-valued condensate. (Again
$h(\alpha/2)$ appears,  the winding 1 loop in $ U(1)_{S_{\hat{x}}}$, as our minimal single-valued choice in  constructing  $U(\varphi, \alpha)$.)  $U(\varphi, \alpha)$ interpolates between
$ h_o^{-1}$ at $\varphi = 0$ and $ h^{-1}(\alpha)$ at $\varphi =
2\pi$. It is thus the fundamental antimonopole in the model, which
contains monopoles and antimonopoles of even winding in $ U(1)_{S_{\hat{x}}}$
only. Note that the inverse twisted Alice loop, with twisted Wilson
line $ U^{-1}(\varphi, \alpha)$, again generates the fundamental monopole.

\section{\label{model:nontop}A Nontopologically Alice string}

A nontopologically Alice string arises when a Higgs $\phi$, transforming
in the adjoint representation under $G=SO(6)$, acquires the vev $\vev
= {\rm diag}(1^3,-1^3)$. As discussed in \cite{stringzm}, this
condensate leaves unbroken an $SO(3)\times SO(3)$ subgroup of $SO(6)$
and a discrete $Z_2$ transformation $h_1=-1\mkern-6mu 1_6$, so
$H=SO(3)\times SO(3)\times Z_2$.  Here $\pi_o(H)=Z_2$ and $\pi_1(H) =
Z_2 \times Z_2$, so topological strings and monopoles form, with
monopoles and antimonopoles topologically identified.  As noted in
\cite{topalice}, strings in this model can have algebraic Alice
behavior, for $U(2\pi) = h_o = {\rm diag} (1^2, (-1)^4) = -
R_{12}(\pi) $, which makes generators $T_{13}$ and $T_{23}$ of
rotations in the $13$- and $23$-planes double-valued. Yet that Alice
behavior fails our topological criterion.  Taking as our nontrivial
monopole loop $h(\alpha) = R_{13}\ (\alpha)$, with monopole charge
(1,0), we find $\tilde{h} (\alpha) = h_o\ h(\alpha)\ h_o^{-1} = h^{-1}
\, (\alpha)\ $, since $T_{13} \rightarrow -\ T_{13}$. Thus the
(1,0) monopole transforms into its antimonopole on traversing the
string. {\bf However,} as monopoles and antimonopoles are identified,
that transformation is nontopological. This string's Alice behavior is
thus nontopological; it can be deformed away by deforming $U(2\pi)$ to the
topologically equivalent value $h_1$, with no algebraic Alice behavior.

We might still hope to construct a (1,0) monopole as a twisted string
loop, taking for our string the algebraic, but nontopologically Alice
string $U(\varphi) = R_{34}(\varphi/2) \ R_{56}(\varphi/2)$, with
algebraic Alice flux $U(2\pi) = h_o = - R_{12}(\pi) $ as above. From
Eq.~(\ref{twistWil}),the twisted Wilson line
$$ U(\varphi, \alpha) = h^{-1}(\alpha)\ U(\varphi)\ h(\alpha)\ h_o^{-1}\ \
$$
generates an Alice loop with single-valued condensate.  $U(\varphi,
\alpha)$ interpolates between $ h_o^{-1}$ at $\varphi = 0$ and $
h^{-2}(\alpha)$ at $\varphi = 2\pi$. This is a loop in $H$ of winding
(2,0); however, winding (2,0) loops are deformable to the identity in $H$, so
this twisted nontopologically Alice loop fails to carry topological
monopole charge.

Recall that, in building singlevalued twisted Alice loops, we required
$ U(\varphi, \alpha)$ to be singlevalued in $\alpha$; we thus
identified $h(\alpha)$ as a loop in $H$. Strictly, we do not need
$h(\alpha)$ to be a loop; all we need is
\beq h^{-1}(2\pi)\ U(\varphi)\ h(2\pi) = U(\varphi)\  .\label{singval}\eeq
We might still hope to build the  fundamental (1,0) monopole as a twisted Alice loop, exploiting this freedom in $h(\alpha)$. Were the twisted loop
$$ U(\varphi, \alpha) = h^{-1}(\alpha/2)\ U(\varphi)\ h(\alpha/2)\ h_o^{-1}\ \
$$
single-valued, it would carry fundamental (-1,0) monopole charge, as it  interpolates between
$ h_o^{-1}$ at $\varphi = 0$ and the nontrivial (-1,0) antimonopole $ h^{-1}(\alpha)$ at $\varphi =
2\pi$. However, this twisted loop candidate is not single-valued; it obeys instead
$ h^{-1}(2\pi)\ U(\varphi)\ h(2\pi) = U^{-1}(\varphi)\ .$ We
thus cannot build a fundamental (1,0) monopole as a twisted Alice loop in
this model, where Alice behavior is nontopological and monopole
charge is $Z_2 \times Z_2$.

This possibility to construct $U(\varphi,\alpha)$ single-valued in
$\alpha$, {\bf without} forcing $h(\alpha)$ to be a loop, always
merits investigating. Indeed, in \cite{skyrme, global}, one of us
constructed what is essentially the fundamental monopole in this
model, by exploiting exactly such an accidental algebraic
singlevaluedness. That construction (most clearly in section IIIA of
\cite{global}, taking $F(r), \varphi$ as the spherical coordinates
$\theta, \varphi$ at spatial infinity),is quite similar to the
twisting constructions here. However, it describes a fundamentally
point-like defect, and cannot be interpreted as a twisted loop.

\section{\label{model:su3}A Topologically Alice Loop Carrying only Deposited Monopole Charge }

We consider a slightly modified canonical Alice string. Take $G$ to be
$SU(3)$, with Higgs $\phi$ transforming according to $\phi \rightarrow
g\, \phi\, g^T$ (giving fermions in this model a Majorana mass).
When $\phi$ develops the vev
$\vev = {\rm diag}\ (1,1,-2)\ \ ,
$
$SU(3)$ breaks to the residual symmetry $H = O(2)$,
identical to that of the canonical Schwarz Alice string.
Again we have $\pi_o (H) = Z_2$ and $\pi_1(H) = Z$,
with topological strings and monopoles. We have the same Alice string
as in the canonical case, making the $O(2)$ generator $T_z$
double-valued. This Alice behavior is again topological, as our topological
criterion, that $\pi_o(H)$ acts nontrivially on $\pi_1(H)$, depends
only on the unbroken symmetry group $H$.

Where we deviate from the canonical Alice string model is in the
identification of twisted Alice loops as monopoles. Here, by the exact
sequence for $\pi_2 (G/H)$, topological monopoles are associated with
nontrivial loops in $O(2)$ which can be unwound in $G$, here
$SU(3)$. {\bf All} nontrivial loops in $O(2)$ can be unwound in
$SU(3)$; thus the fundamental monopole in this model  has a loop
in $O(2)$ of winding 1.

We now construct a monopole as a twisted Alice loop. We take
$U(\varphi) = R_x\, (\varphi/2)$, as in the canonical Alice case, and
$h(\alpha) = R_z(\alpha)$, a loop of winding 1. From
Eq.~(\ref{twistWil}), the twisted Wilson line
$$ U(\varphi, \alpha) = h^{-1}(\alpha)\ U(\varphi)\ h(\alpha)\ h_o^{-1}\ \
$$ generates a twisted Alice loop with single-valued condensate.
$U(\varphi, \alpha)$ interpolates between $ h_o^{-1}$ at $\varphi = 0$
and $ h^{-2}(\alpha)$ at $\varphi = 2\pi$. This twisted Alice loop
carries monopole charge of $-2$, which while nontrivial is {\bf not}
the fundamental antimonopole in this model. (Similarly, the inverse
twisted Alice loop, with Wilson line $ U^{-1}(\varphi, \alpha)$,
carries monopole charge +2).

Again, we might still hope to build a fundamental monopole as a
twisted Alice loop, by allowing $h(\alpha)$ above to be not a loop,
but a curve obeying Eq.~(\ref{singval})
This looser
constraint still guarantees singlevaluedness in $\alpha$ of $
U(\varphi, \alpha)$.  Indeed, were the twisted loop
$$ U(\varphi, \alpha) = h^{-1}(\alpha/2)\ U(\varphi)\ h(\alpha/2)\ h_o^{-1}\ \
$$ single-valued in $\alpha$, with $h(\alpha) = R_z(\alpha)$ as above,
it would carry fundamental antimonopole charge. This is because it
interpolates between $ h_o^{-1}$ at $\varphi = 0$ and the winding $-1$
loop $ h^{-1}(\alpha)$ at $\varphi = 2\pi$. However, this twisted loop
candidate is not single-valued in $\alpha$; it obeys instead
$ U(\varphi, 2\pi) = R_x^{-1} (\varphi) \ U(\varphi, 0)\  .
$ We thus cannot build a fundamental monopole as a twisted Alice loop
in this model. Instead twisted Alice loops carry only the monopole
charge which topological arguments ensure they must carry: because
monopoles scatter into antimonopoles on transiting Alice loops, Alice
loops {\bf must} support deposited monopole charge, which arises in
units of 2. Our twisting construction creates twisted Alice loops
supporting exactly that deposited charge.

\begin{acknowledgments}
Early stages of this work were supported by NSF grant PHY-9631182 and
by the University Research Committee of Emory University.
\end{acknowledgments}

\end{document}